Tomasz Toliński

# Thermoelectric Power of the CeCoAl$_4$ Antiferromagnet – Predominance of the Crystal Electric Field

Institute of Molecular Physics, Polish Academy of Sciences,

M. Smoluchowskiego 17, 60-179 Poznań, Poland

Corresponding author:

e-mail tomtol@ifmpan.poznan.pl

**Abstract**

We report on the thermoelectric power $S(T)$, electrical resistivity and thermal conductivity measurements on the antiferromagnetic CeCoAl$_4$ compound. The Néel temperature of CeCoAl$_4$ is $T_N$ = 13.5 K and a metamagnetic-like transition occurs in the magnetic field of about 7.5 T. We show that the magnetic contribution to the thermoelectric power, $S_f(T)$, exhibits a large peak due to the crystal electric field (CEF) at about 240 K and a small anomaly around the ordering temperature. Surprisingly, the CEF related $S_f(T)$ peak is close to the upper CEF excitation energy $\Delta_2$ =202 K, in contrast to the usual case of the peak position being a fraction of the real CEF excitation. The applicability of different theoretical and phenomenological models describing the temperature dependence of the Seebeck effect $S_f(T)$ has been tested and finally a calculation is proposed, which incorporates a full CEF levels scheme. For the nonmagnetic reference compound, LaCoAl$_4$, the thermoelectric power measurements provide an evidence of the phonon drag contribution, which is not present or masked in the case of CeCoAl$_4$.

# 1 Introduction

The CeCoAl$_4$ compound poses a fascinating topic for in-depth studies due to its magnetic properties, which are governed solely by the Ce atoms. Co atoms are non-magnetic in CeCoAl$_4$. This compound is characterized by antiferromagnetic (AFM) order below $T_N$ = 13.5 K and a metamagnetic-like transition in the magnetic field of about 7.5 T [1-5]. The crystallographic structure is orthorhombic of the LaCoAl$_4$ type (space group Pmma) with PrCoAl$_4$ being the other known example of this structure type [6].

Dhar *et al*. [4] suggested an incommensurate antiferromagnetic order for CeCoAl$_4$ based on magnetic susceptibility, electrical resistivity and specific heat measurements, however neutron diffraction experiments [6] have revealed a collinear antiferromagnetism with a propagation vector **q** = (0, 0.5, 0.5) and the refined magnetic moment equal to 1.29 $\mu_B$/Ce atom. In other studies Koterlin *et al*. [7] postulated that CeCoAl$_4$ represents a magnetic Kondo lattice. This suggestion has been based on the analysis of the magnetic contribution to resistivity and on the magnetic susceptibility measurements.

In our previous studies [8] we have shown that a small addition of Cu to the CeCoAl$_4$ compound stabilizes the main phase by a reduction of the amount of the impurity phases, mainly CeAl$_2$. For CeCo$_{0.9}$Cu$_{0.1}$Al$_4$ the secondary phases (about 3%) are created by CeAl$_2$ and CeCuAl$_3$. For the Cu-stabilized CeCo$_{0.9}$Cu$_{0.1}$Al$_4$ compound and below the Néel temperature $T_N$ = 13.5 K we have observed an antiferromagnetic order with the propagation vector **q** = (0, 1/2, 1/2). The magnetic moment of 0.76 $\mu_B$/Ce was reduced compared to the theoretical Ce moment and the previous results for CeCoAl$_4$. We have also indicated on the importance of the crystal electric field (CEF) effects [9]. Our complementary studies using the inelastic neutron scattering, magnetic susceptibility and specific heat measurements provided the CEF parameters and the energy levels scheme composed of three doublets (0–4.04–17.4) meV. In the ground state we have found a mixing of the $|1/2\rangle$, $|3/2\rangle$, and $|5/2\rangle$ states.

In this paper we discuss the transport properties of CeCoAl$_4$ including the thermoelectric power $S(T)$, thermal conductivity $\lambda(T)$, and electrical resistivity $\rho(T)$. A special emphasis is put on the verification of different models to describe the temperature dependence of the Seebeck effect $S(T)$ and a new solution for including CEF in the analysis of the thermoelectric power is proposed.

## 2 Experimental

The polycrystalline CeCoAl$_4$ compound was synthesized by induction melting under an argon atmosphere. Stoichiometric amounts of the constituent elements were used. The sample was inverted and melted several times to ensure a better homogeneity.

The crystal structure and the quality of the sample was controlled using the X-ray diffraction technique [8,9].

The thermoelectric power, electrical resistivity and thermal conductivity were measured using the thermal transport option (TTO) of the commercial PPMS device (Quantum Design). Four-probe mode was employed in all the measurements. The samples are cut into a bar of the approximate dimensions 1mm×1mm×8mm.

## 3 Results and Discussion

Figure 1 presents the results of measurements of the thermoelectric power $S$, thermal conductivity $\lambda$, and the electrical resistivity $\rho$ for the LaCoAl$_4$ compound measured in a wide temperature range 1.9-400K. This compound has been used as the nonmagnetic analog to extract the f-electrons contribution of the CeCoAl$_4$ compound. However, LaCoAl$_4$ in itself shows intriguing features. In low temperatures the $S(T)$ dependence shows a flat bump, which is better visible if one plots $S/T$ vs. $T^2$. It shows a linear dependence in the temperatures range ~25-50 K. Such a linear behavior is typical of phonon-drag contribution if it occurs in the range $\Theta/10$-$\Theta/5$, where $\Theta$ is the Debye temperature. As for LaCoAl$_4$ we have got $\Theta$~320K, the condition is well fulfilled.

Figure 1 presents additionally the temperature dependence of the thermal conductivity with the separated phonon and electronic contributions. The phonon part has been determined assuming that $\lambda = \lambda_{el} + \lambda_{ph}$ and the electronic contribution is connected with the resistivity via the Wiedemann-Franz law $\lambda_{el} = L_0 T/\rho$ ($L_0 = 2.45\times10^{-8}$ W$\Omega$K$^{-2}$ is the Lorentz number). It appears that the phonon contribution predominates below about 200 K and is close to $\lambda_{el}$ but still slightly larger above 200 K.

Figure 2 shows similar analysis for the CeCoAl$_4$ compound. One can notice that the residual resistivity in our case is similar to the result of Koterlyn et al. [7]. The values of $\rho_0$ are: 70, 20, 45 $\mu\Omega$ cm for LaCoAl$_4$ and 200, 120, 100 $\mu\Omega$ cm for CeCoAl$_4$ for the Dhar et al. [4], Koterlyn et al. [7], and our studies, respectively. Differences between these results can result from a different amount of microcracks, which was also mentioned in Ref. [4]. Moreover, in our neutron diffraction measurements [8] we indicated that a small amount of impurities is possible, up to 3%, e.g. CeAl$_2$ type but treating it as a parallel resistance shows in simple calculations that it can modify the calculation of $\lambda_{el}$ by only a few percent and towards lower values, i.e. even strengthening the conclusion that $\lambda_{el}$ is lower than $\lambda_{ph}$ in the whole studied temperature range.

From Fig. 2 it is also concluded that the phonon-drag contribution is no longer detectable, which is probably due to the dominance of the magnetic interactions at the low temperatures region. A drop in $S(T)$ is well visible close to $T_N$.

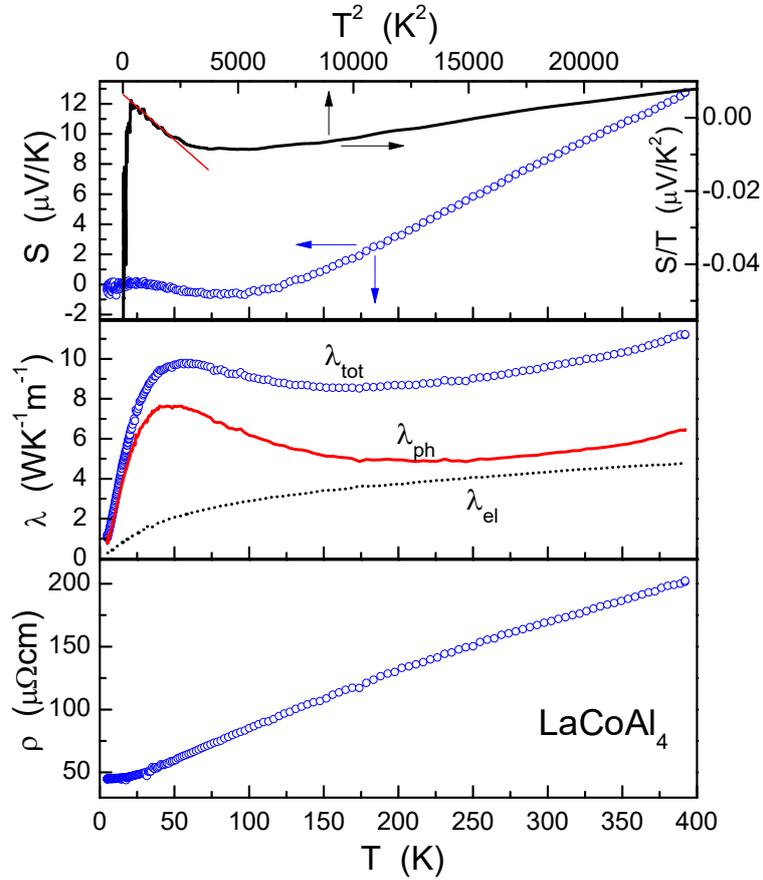

**Fig. 1** Thermoelectric power $S$, thermal conductivity $\lambda$, and the electrical resistivity $\rho$ for the LaCoAl$_4$ compound. Thermoelectric power is shown both as $S$ vs. $T$ and $S/T$ vs. $T^2$. Solid line above the bump for $S/T(T^2)$ is a linear fit characteristic of the phonon-drag contribution. The total thermal conductivity is separated to the phonon and electronic contribution via the Wiedemann-Franz law (see text)

To estimate the contribution of the f states to the thermoelectric power of CeCoAl$_4$ we apply the Northeim-Gorter rule: $S\rho = S_f \rho_f + S_{nmag} \rho_{nmag}$, where $\rho_f$ is obtained by subtraction the resistivity of the La-based analogue sample, i.e. $\rho_{nmag}$. The extracted dependence of $S_f$ on temperature is plotted in Fig. 2. It exhibits a large peak at about 240 K and a small anomaly around the ordering temperature. The peak at 240 K is clearly related to the crystal electric field effects.

Interestingly, the f-electrons contribution to the resistivity (Fig. 2) reveals two-peak structure with the peaks at about 220 K and 120 K. We ascribe both to the CEF levels and the smearing out of the lower peak for thermoelectric power may be due to the different mechanisms governing $S(T)$ behavior compared to $\rho_f$ including only the electronic contribution. The sharp anomaly in $\rho_f(T)$ at 13.5 K corresponds to the Néel temperature.

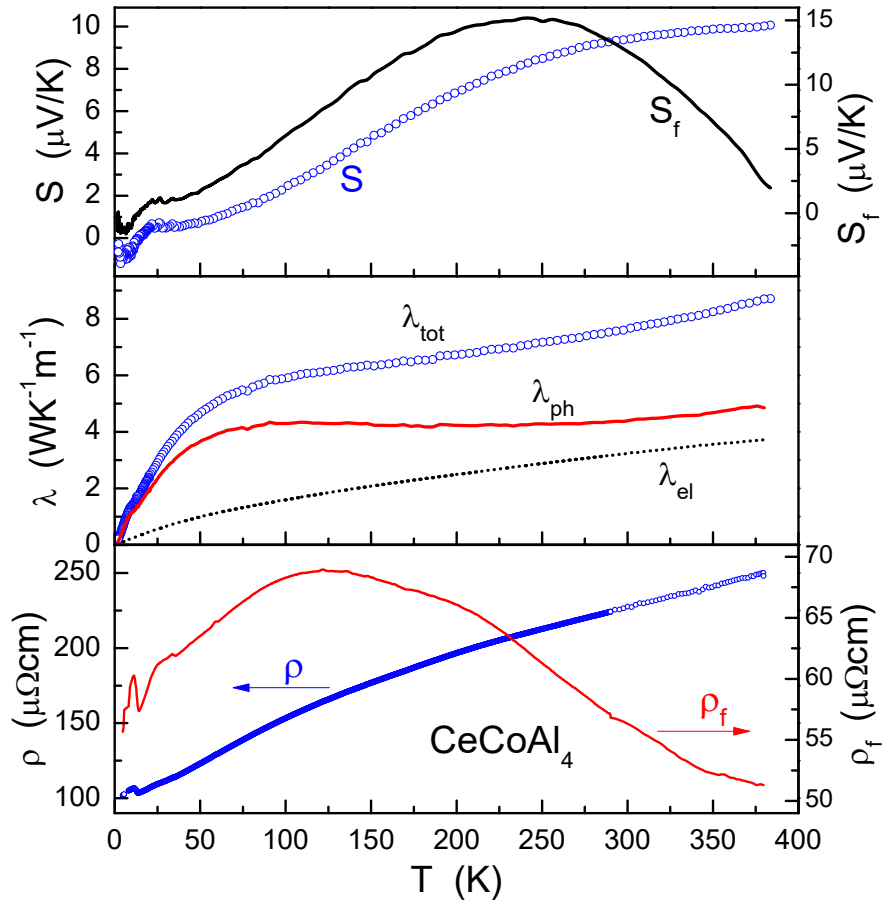

**Fig. 2** Thermoelectric power $S$, thermal conductivity $\lambda$, and the electrical resistivity $\rho$ for the CeCoAl$_4$ compound. $S_f$ is the f–electrons contribution to the thermoelectric power obtained from the Northeim-Gorter rule. The total thermal conductivity is separated to the phonon and electronic contribution via the Wiedemann-Franz law (see text). $\rho_f$ is the f–electrons contribution to resistivity derived by subtraction the resistivity of LaCoAl$_4$

Our previous inelastic neutron scattering (INS) experiments [9] provided the CEF energy levels scheme (0 – 4.04 – 17.4) meV, which corresponds to (0 – 47 – 202) K. The higher level excitation is near the maximum in $S_f(T)$. This is surprising as it is known that in the f-electron systems in the presence of the hybridization of the f electrons with the conduction band, the Kondo scattering on the excited CEF levels leads to a peak in $S_f(T)$ at $T_{max} \approx (0.3-0.6)\Delta_{CEF}$. A possible explanation of this untypical behavior of the CeCoAl$_4$ compound is our previous observation of the behavior of the quasielastic contribution in the INS spectra. The inelastic broadening disappears below the Néel temperature and its linewidth $\Gamma/2$ is below $T_N$, therefore it suggests that the Kondo screening is not relevant in CeCoAl$_4$. Nevertheless, we test below the main theoretical and semi-phenomenological models describing the temperature dependence of the thermoelectric power in the f-electron systems, including models based on the Kondo effect.

**3.1 Peschel and Fulde model**

The calculations of Peschel and Fulde [10] concern metals with impurities affected by the crystal-field split energy levels, hence the problem is of different type than the Kondo problem. Therefore, we apply it to the present case of CeCoAl$_4$ in spite of the fact that Ce does not state an impurity in this compound but creates a lattice. Peschel and Fulde considered two nonmagnetic singlets separated by an energy $\Delta_{CEF}$ and derived the relation:

$$S_1(T) \propto \frac{\Delta_{CEF}}{2T} th \frac{\Delta_{CEF}}{2T} \left[1 + \frac{\Delta_{CEF}}{2\pi T} \text{Im} \psi^{(1)}\left(\frac{i\Delta_{CEF}}{2\pi T}\right)\right], \tag{1}$$

where $\psi^{(1)}$ is a trigamma-function.

The plot according to the formula $S_f(T) = a \cdot T + b \cdot S_1(T)$, where the first term is the nonmagnetic contribution, is included in Fig. 3 for $a = -0.09$ µV/K$^2$ and $b = 45$. The position of the maximum is well reproduced only for $\Delta_{CEF} = 1120$ K, which is over five times the value of the upper excitation derived from our INS experiments (202 K). It could be improved by using a real levels degeneracies, i.e. three doublets instead of two singlets.

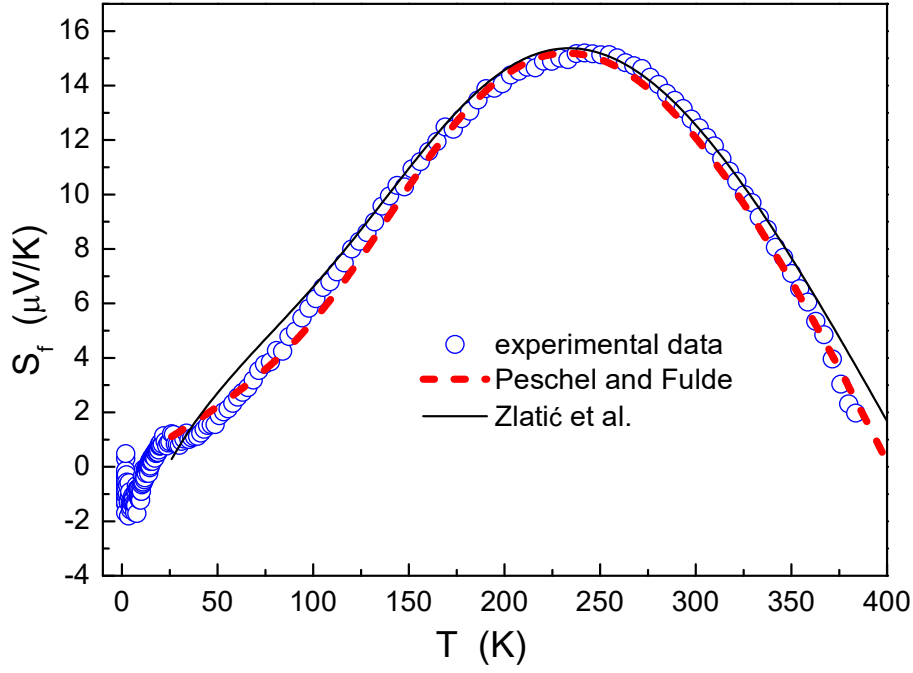

**Fig. 3** Thermoelectric power $S_f(T) = a \cdot T + b \cdot S_1(T)$ of CeCoAl$_4$ fitted with: red dotted line - Peschel and Fulde model (Eq.(1)), solid line – Zlatić-Coqblin-Schrieffer model (Eq.(3))

### 3.2 Zlatić-Coqblin-Schrieffer model

Alternatively, we have employed the calculations proposed by Zlatić et al. for the case of two CEF levels [11]. They have shown that the Coqblin-Schrieffer model in connection with the third-order renormalized perturbation expansion can well describe the influence of CEF on the thermoelectric power in the heavy-fermion systems and can lead to a reduction of the Kondo temperature $T_K$. The renormalized coupling constant is given by:

$$\exp\left(\frac{2D_0}{J}\right) = \left(\frac{T_K}{D}\right)^m \left(\frac{T_K + \Delta_{CEF}}{D + \Delta_{CEF}}\right)^M, \qquad (2)$$

where $2D_0$ is the width of the conduction band, $D = T$ and $\Delta_{CEF}$ is the separation between the ground and the excited state, which are $m$ and M-fold degenerate, respectively.

The thermoelectric power of the two level system (doublet-quartet) is described by the formula:

$$S_1(T) = \frac{k_B}{|e|} \frac{S_\Delta}{R_\Delta} G_1(\Delta_{CEF}, 0) \qquad (3)$$

with the respective terms defined in details in Ref. [11]. Assuming $D_0 = 11610$ K (1 eV) and a formula $S_f(T) = a \cdot T + b \cdot S_1(T)$, we get the solid line in Fig. 3 for $V_0 = -0.14 \times 2D_0$, $T_K = 1$ K, $a = -6.2 \times 10^{-8}$ μV/K$^2$, $b = 3.55$ and $\Delta_{CEF} = 1240$ K. $V_0$ is a coupling constant related to the potential scattering (for details see [11]). Again, like in sec. 3.1, the CEF splitting necessary for reproducing the $S(T)$ peak position is much higher than the result of INS studies. Additionally, a Kondo temperature is involved, which is not adequate for the case of CeCoAl$_4$.

### 3.3 Two band model

The thermoelectric power calculated from the linearized Boltzmann equation [12-16] can be written as:

$$S(T) = -\frac{1}{|e|T} \frac{I_1}{I_2}, \qquad (4)$$

where the integrals

$$I_n = \int_{-\infty}^{+\infty} E^n \frac{dF}{dE} \tau dE, \quad n = 0,1, \qquad (5)$$

with $F(E)$ being the Fermi function.

In the degenerate limit and using the Sommerfeld expansion provides the Mott expression [12,13]:

$$S(T) = \frac{\pi^2 k_B^2 T}{3|e|} \left(\frac{\partial \ln\sigma}{\partial E}\right)_{E_F}, \qquad (6)$$

where $\sigma$ is the electrical conductivity. In the free-electron approximation one obtains for the diffusion thermoelectric power:

$$S(T) = aT. \qquad (7)$$

To include, apart from the diffusion thermoelectric power, the contribution of the f states, we have carried out an analysis of the measured thermoelectric power basing on the model assuming scattering of electrons from the wide conduction band into a narrow f band approximated by a density of states of the Lorentzian shape [14-16]:

$$N_f(E) \propto W_f / \left( (E - E_f)^2 + W_f^2 \right). \tag{8}$$

With the further assumption of the relaxation time $\tau \sim N_f(E)^{-1}$ and employing Eq. (4), thermoelectric power is expressed by the formula [14-16]:

$$S_f(T) = \frac{2}{3} \frac{k_B}{|e|} \frac{\pi^2 E_f T}{(\pi^2/3)T^2 + E_f^2 + W_f^2}, \tag{9}$$

where $E_f$ and $W_f$ are the position and width of the $f$-band in Kelvins and $S_f$ denotes the $f$ contribution to thermoelectric power. Eq. (9) has been widely used [14-21] for analysis of thermoelectric power in Ce intermetallics providing an estimation of the f-states position in respect to the Fermi level, however it does not include the effect of the CEF splitting.

The derivation of Eq. (9) based on Eq. (4) is simple for a single DOS peak but, in principle, for a CEF scheme composed of three doublets a derivation for a sum of three DOS peaks would be required. However, we have recently shown that $W_f$ can be treated as dominated by the CEF split, i.e. the width of the DOS peak being of the order of the ground state split [17], $W_f = \pi T_{CEF}/N_f$, where $N_f$ is the orbital degeneracy 2J + 1. This approximation has been successfully applied in various Ce-based systems [17,22-24].

As the shape of $S_f(T)$ can suffer from errors due to the approximation of the Northeim-Gorter rule it is advantageous to employ the position of the thermoelectric power maximum as a reasonable characteristic of the system. Taking derivative of Eq. (9) in respect to temperature with the condition $\partial S_f/\partial T = 0$ leads to the relation:

$$W_f = 1.74 \times 10^{-8} \sqrt{1.08 \times 10^{16} T^2 - 3.29 \times 10^{15} E_f^2} \tag{10}$$

with all magnitudes in Kelvins. Figure 4 illustrates the function $W_f(E_f)$ for $T = T_{max} = 240$ K. Assuming $N_f = 6$ and $T_{CEF} \approx \Delta_{CEF} \approx 202$ K we expect $W_f = 106$ K, which implies, according to Eq. (9) and Fig. 4, $E_f = 422$ K.

As mentioned above, it is problematic to use Eq. (4) for the case of three DOS peaks with CEF separations equal to $\Delta_1$ and $\Delta_2$. Therefore, to verify the applicability of the estimations done by Eq. (9) and Eq. (10) we refer to the approximation used by Freimuth [25,26]. He used the special case of the thermoelectric power formula:

$$S(T) = \frac{\pi^2 k_B T}{3|e|} \left(\frac{\partial \ln D(E)}{\partial E}\right)_{E_F} \qquad (11)$$

expressed by the derivative of DOS at $E_F$. Such approximation neglects the role of the Fermi distribution responsible for the temperature dependence in formula (5). The assumption in Ref. [25] was to make the width of the DOS peak as well as the Kondo temperature dependent on temperature. To reduce the number of free parameters we restrict this assumption only to the width dependence of the form $W_f = v \cdot \exp(-v/T)$. Hence, the sum of the three DOS peaks can be written as:

$$D(E) = \sum_{i=0}^{2} \frac{v_i \exp(-v_i/T)}{\left(E - E_f + \Delta_i\right)^2 + v_i^2 \exp(-2v_i/T)}, \qquad (12)$$

which after application of Eq. (12) leads to:

$$S_1(T) = \frac{2\pi^2 k_B T}{3|e|} \left\{ \sum_{i=0}^{2} \frac{v_i(E_f + \Delta_i)\exp(-v_i/T)}{\left[\left(E_f + \Delta_i\right)^2 + v_i^2 \exp(-2v_i/T)\right]^2} \right\} / \left\{ \sum_{i=0}^{2} \frac{v_i \exp(-v_i/T)}{\left(E_f + \Delta_i\right)^2 + v_i^2 \exp(-2v_i/T)} \right\} \qquad (13)$$

where the substitution $(E_f - E_F) \rightarrow E_f$ was done, i.e. $E_f$ is considered in respect to the Fermi level. Figure 5 shows a fit of the CeCoAl$_4$ thermoelectric power by Eq. (13) adding the usual linear term, i.e. $S_f(T) = a \cdot T + S_1(T)$, fixing only the CEF energies with the INS values: $\Delta_0 = 0$ K, $\Delta_1 = 47$ K, and $\Delta_2 = 202$ K. As a result the following values of the remaining parameters have been obtained: $v_1 = 486$ K, $v_2 = 353$ K, $v_3 = 372$ K, $a = -1.07 \times 10^{-6}$ μV/K$^2$, and $E_f = 435$ K - the last being in a very good agreement with the estimation performed by Eq. (11) ($E_f = 422$ K).

It is interesting to demonstrate what the DOS structure looks like for the methodology based on Eq. (9) and Eq. (13). It is drawn as the left-bottom axes in Fig. 4. For Eq. (13) the three components are shown and $T = T_{max} = 240$ K is used in the calculation. One can notice that the width of the single DOS peak of Eq. (9) is roughly of the order of the overall CEF split. However, in spite of the fact that the model of Eq. (12) is formally less correct than that based directly on Eqs. (5-6), it provides a reasonable values of the parameters. Even setting in Eq. (13)

$\Delta_1$ and $\Delta_2$ as free parameters has given physically correct values of these splits (55 K and 90 K, respectively).

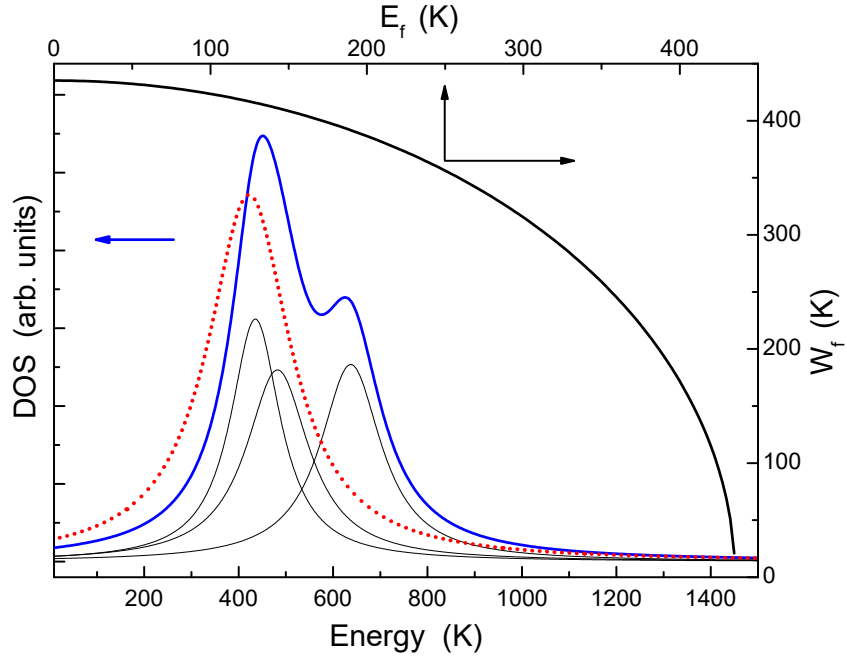

**Fig. 4** Right and top axes: $W_f$ vs. $E_f$ for $T = T_{max} = 240$ K plotted according to Eq. (10). Left and bottom axes: density of states – dotted line corresponds to Eq. (9) and the thick solid lines corresponds to Eq. (13). The thin solid lines are the three components of Eq. (13)

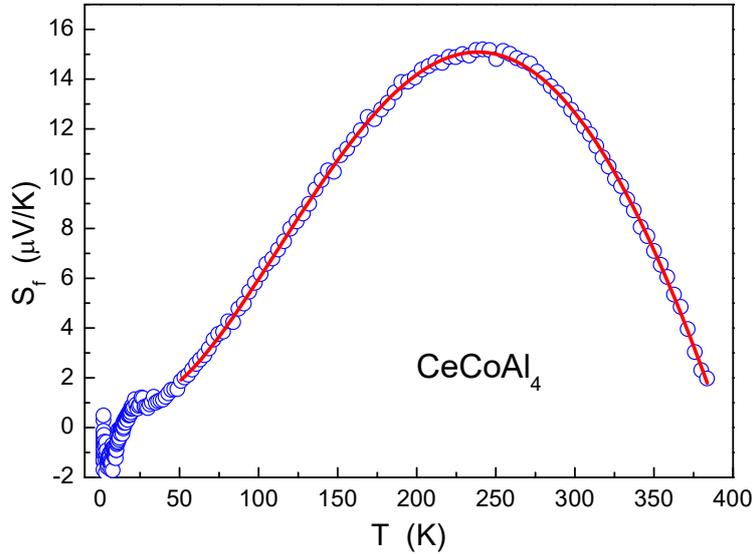

**Fig. 5** f-electrons contribution to the thermoelectric power of CeCoAl$_4$ fitted by Eq. (13): $S_f(T) = a \cdot T + S_1(T)$. The CEF energies are fixed with the previous result of the inelastic neutron studies: $\Delta_0 = 0$ K, $\Delta_1 = 47$ K, and $\Delta_2 = 202$ K

## 4 Conclusions

The CeCoAl$_4$ compound, which we have studied previously in details by inelastic neutron scattering, neutron diffraction, as well as magnetic and specific heat methods has now been used to test applicability of different models of the thermoelectric power. The key challenge was to incorporate the predominance of the crystal field in the theoretical description of the thermoelectric power $S_f(T)$. The well-established theoretical models are usually based on the assumption of a single split of the CEF levels, which does not enable refinement of the thermoelectric power of CeCoAl$_4$, because this compound exhibits a split into three CEF doublets. In the present calculation we have assumed a multi-peak density of states with a structure corresponding to the real CEF levels scheme. The temperature dependence has been included making the width of the DOS peaks temperature dependent. It appeared to fit well the $S_f(T)$ dependence, with physically correct parameters. Alternatively, the use of a single DOS peak but with temperature dependence resulting directly from the Fermi function provides a good estimate of the overall CEF splitting but cannot reveal the detailed levels scheme.

The usual studies of the nonmagnetic reference compound, LaCoAl$_4$ in our case, revealed a possible contribution of the phonon drag in the temperature range ~25-50 K of the thermoelectric power.

The f-electrons contribution to the resistivity has shown a presence of two peaks, at about 220 K and 120 K, which can be ascribed to the CEF levels.

**Acknowledgements** Karol Synoradzki is acknowledged for the experimental assistance.

## References

1. L. D. Tung, N. P. Thuy, J. J. M. Franse, P. E. Brommer, J. H. P. Colpa, J. C. P. Klaasse, F. R. de Boer, A. A. Menovsky, K. H. J. Buschow, J. Alloys Compd. **281**, 108 (1998)
2. L. D. Tung, J. J. M. Franse, K. H. J. Buschow, P. E. Brommer, J. H. P. Colpa, J. C. P. Klaasse, A. A. Menovsky, J. Magn. Magn. Mater. **177-181**, 477 (1998)
3. O. Moze, L. D. Tung, J. J. M. Franse, K. H. J. Buschow, J. Alloys Compd. **256**, 45 (1997)
4. S. K. Dhar, B. Rama, S. Ramakrishnan, Phys. Rev. B **52**, 4284 (1995)
5. A. Schenck, F. N. Gygax, P. Schobinger-Papamantellos, L. D. Tung, Phys. Rev. B **71**, 214411 (2005)

6. P. Schobinger-Papamantellos, C. Wilkinson, C. Ritter, L. D. Tung, K. H. J. Buschow, O. Moze, J. Phys. Condens. Matter **16,** 6569 (2004)
7. M. D. Koterlin, B. S. Morokivskii, R. R. Kutayanskii, N. G. Babic, N. I. Zakharenko, Phys. Solid State **39,** 456 (1997)
8. T. Toliński, A. Hoser, N. Stüßer, A. Kowalczyk, D. Kaczorowski, J. optoelectronics and advanced materials **14,** 90 (2012)
9. T. Toliński, K. Synoradzki, A. Hoser, S. Rols, J. Magn. Magn. Mater. **345,** 243 (2013).
10. I. Peschel, P. Fulde, Z. Physik **238,** 99 (1970)
11. V. Zlatić, B. Horvatić, I. Milat, B. Coqblin, G. Czycholl, C. Grenzebach, Phys. Rev B **68,** 104432 (2003)
12. F. J. Blatt, P. A. Schroeder, C. L. Foiles, D. Grieg, Thermoelectric Power of Metals (Plenum, New York, 1976)
13. N. F. Mott, H. Jones, The Theory of the Properties of Metals and Alloys, Clarendon Press, Oxford (1936)
14. M. D. Koterlyn, R. I. Yasnitskii, G. M. Koterlyn, B. S. Morokhivskii, J. Alloys Compd. **348,** 52 (2003)
15. M. D. Koterlyn, O. Babych, G. M. Koterlyn, J. Alloys Compd. **325,** 6 (2001)
16. U. Gottwick, K. Gloos, S. Horn, F. Steglich, N. Grewe, J. Magn. Magn. Mater. **47-48,** 536 (1985)
17. T. Toliński, V. Zlatić, A. Kowalczyk, J. Alloys Compd. **490,** 15 (2010)
18. T. Toliński, Eur. Phys. J. B **84,** 177 (2011)
19. Sankararao Yadam, Durgesh Singh, D. Venkateshwarlu, Mohan Kumar Gangrade, S. Shanmukharao Samatham, V. Ganesan, Phys. Status Solidi B **252,** 502 (2015)
20. A. P. Pikul, D. Kaczorowski, Z. Bukowski, K. Gofryk, U. Burkhardt, Yu. Grin, F. Steglich, Phys. Rev B **73,** 092406 (2006)
21. K. Gofryk, J.-C. Griveau, P. S. Riseborough, T. Durakiewicz, Phys. Rev B **94,** 195117 (2016)
22. A.K. Bashir, M. B. Tchoula Tchokonté, D. Britz, A. M. Strydom, D. Kaczorowski, J. Phys. Chem. Solids **106**, 44 (2017)
23. M. B. Tchoula Tchokonté, A. K. Bashir, A. M. Strydom, T. D. Doyle, D. Kaczorowski, J. Alloys Compd. **717**, 333 (2017)
24. M. Falkowski, A. Kowalczyk, Intermetallics **20,** 173 (2012)
25. A. Freimuth, J. Magn. Magn. Mater. **68,** 28 (1987)
26. C. S. Garde, J. Ray, Phys. Rev B **51,** 2960 (1995)